# SUSPENSION OF NANOPARTICLES IN SU-8 AND CHARACTERIZATION OF NANOCOMPOSITE POLYMERS

H.C. Chiamori[1], J.W. Brown[1], E.V. Adhiprakasha[1*], E.T. Hantsoo[1**], J.B. Straalsund[1], N.A. Melosh[2], B.L. Pruitt[1]

[1]Department of Mechanical Engineering, Stanford University, CA, USA
[2]Department of Materials Science and Engineering, Stanford University, CA, USA

**ABSTRACT**

Gold nanospheres, single wall carbon nanotubes (SWNT), and diamondoids [1] were physically incorporated into the negative photoresist SU-8 [2]. The mixtures were spin cast onto silicon or aluminum coated silicon wafers. ASTM standard D638 tensile specimens were lithographically patterned in the materials and then released from the substrate using Microchem's Omnicoat or an anodic metal dissolution process. The residual stresses, elastic moduli, and viscosity effects of the nanocomposites were measured and compared to control specimens of SU-8. Resistivity measurements of SU-8/SWNT nanocomposites were also investigated. We found the effective modulus and viscosity of the SU-8 test specimens decreases with the addition of diamantane and SWNTs. Additionally, the SU-8/SWNT nanocomposites showed changes in resistivity with increased strain, suggesting a gauge factor for the 1 wt% SU-8/SWNT nanocomposite of approximately 2-4.

## 1. INTRODUCTION

Interest in using photosensitive polymers for microscale sensing applications has grown as the need for biological sensors inert to the body has emerged, silicon equipment and fabrication costs have risen, and silicon fabrication steps have become increasingly complicated. Photosensitive polymers are readily available, tough, chemically inert, and mechanically stable which make them ideally suited for robust, low cost, and repeatable manufacturing applications. Current microscale applications for these polymers include high aspect ratio structures [3,4], molds [3,5], microfluidic channels [6], and packaging [7]. Potential uses of modified photoresists for microelectromechanical (MEMS) devices include conductive or piezoresistive transducers, electrical interconnects, and packaging of devices. Several barriers prevent widespread use of lithographically cured polymers for MEMS-based devices, especially the high residual stresses resulting from material processing, optimization of piezoresistive effects of nanoparticles with electrical properties dispersed throughout a polymer matrix, and ease of manufacturing processing methods for the development of these nanocomposite materials.

Prior research on polymer thin films exhibiting piezoresistive effects includes application of conductive inks [8] or deposition of metal layers on top of the structures [9]. Jiguet et al, and Renaud et al successfully dispersed silver particles and silica crystals in SU-8 [10,11]. With the silver powders, the nanocomposite material mixture resulted in a thick paste which could not be spin cast on a wafer [10]. Zhang et al, successful dispersed multi-walled carbon nanotubes in SU-8 [12]. Damean et al, incorporated nickel nanospheres in SU-8 and fabricate magnetically-responsive devices [13].

This research focuses on synthesizing nanocomposite materials by dispersing nanoparticles throughout the polymer matrix and evaluating the resulting effects on bulk material properties. One of our goals is to tailor these "doped" polymer devices for specific applications such as sensing and actuation, biomedical applications, or packaging and interconnects. Our initial work utilizes the epoxy-based negative photoresist SU-8 [2] with formulations allowing thickness ranges of less than 1-um to greater than 200-um. Properties of SU-8 include high optical transparency above 360-nm, sensitivity to near UV radiation (350-400-nm) and resistance to solvents, acids and bases [2]. SU-8 has excellent thermal stability with $T_g$>200deg C after crosslinking the material [5]. We present data on nanocomposite elastic moduli, residual stresses and conductivity as applicable. We are currently working with three types of nanoparticles: gold nanospheres, single-wall carbon nanotubes (SWNT) and diamondoids.

Gold nanoparticles in powder form (diameter range of 100 to 300-nm) were purchased from Sigma-Aldrich. Initial tests used 5-nm gold nanoparticles in colloidal solution from TedPella, although the functionalized colloid gold content was impractical at 12 vol%.

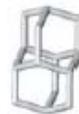

**Figure 1 Structure of Diamantane $C_{14}H_{20}$ [1].**

Single wall carbon nanotubes (SWNT) were purchased from SouthWest NanoTechnologies. The carbon content is specified greater than 90% SWNT compared with other forms and contains the two main semiconducting types of SWNT (6,5) +(7,5). The average diameter of each nanotube is 0.8-nm +/- 0.2-nm [14].

---

* currently at AMD/Spansion and **Nanosolar, CA, USA

The excellent mechanical, electrical, and thermal properties of carbon nanotubes are well-documented, although difficulties with processing the material have hindered their broad scale application [12]. Due to the high surface energy on the molecule, carbon nanotubes tend to agglomerate in solution making dispersion difficult. Functionalizing the carbon nanotubes allows possibility of grafting the modified carbon nanotube to the epoxy rings of SU-8, as demonstrated by Zhang. [12].

Based on percolation theory [15], a strain sensitive conductive path dependent on the mass percent of carbon nanotubes (or metallic nanoparticles) in the system is possible. Functionalized carbon nanotubes have an additional geometric advantage. Higher aspect ratio particles require lower percolation thresholds [16] due to an increased ability to form networks, so lower mass percentages of the material will be required to establish conductivity in the nanocomposite, ultimately lowering the cost of the nanocomposite material.

Although the piezoresistivity of carbon nanotubes is dependent on the orientation of the lattice [17], strain sensing capability has been successfully demonstrated with random orientation of SWNTs [18], which makes integration of carbon nanotubes into thin films possible. Grow et al showed high gauge factors for semiconducting and small-gap semiconducting (up to 400 and 800, respecctively) carbon nanotubes [17]. When a piezoresistive material is subject to stress, both the resistivity and film dimensions change due to the deformation of the material. This sensitivity to stress can be described by the gauge factor, which is the ratio of change in resistance to strain of the piezoresistive film.

$$\frac{\Delta R}{R} = \frac{G_f \cdot \sigma}{E}$$

where $\Delta R / R$ is the relative change in resistance, $G_f$ is the gauge factor of the piezoresistor, E is Young's modulus of specimen, and $\sigma$ is applied stress [17]. With these high gauge factors and strain-sensing ability, once functionalized and integrated into the polymer material, SU-8/CNT nanocomposites may allow for highly sensitive and flexible devices for sensing and actuation.

The diamondoids are supplied by Chevron MolecularDiamond Technologies [1]. Diamondoids are diamond fragments terminated with hydrogen [19,20]. As shown in Figure 1, diamantane ($C_{14}H_{20}$) consists of two face-fused caged subunits of the diamond structure surrounded by hydrogen atoms [19].

Although diamantane are non-conductive, they can be functionalized to bind with SU-8 or other molecules. Additionally, the diamondoids are optically transparent and can be dissolved in solvents compatible with SU-8. The size, rigidity, transparency and ability to functionalize the diamondoids offer the possibility of a uniform dispersion within the SU-8.

## 2. EXPERIMENTAL METHODS

### 2.1. Nanoparticles Preparation

To synthesize the SU-8/gold nanocomposite, SU-8 is dissolved in benzene and gold nanoparticles are functionalized with a solution of benzenethiol in benzene. The functionalized gold particles and SU-8 - benzene are mixed to create a SU-8/gold mixture. As the benzene evaporates, a mixture containing an even dispersion of gold particles in SU-8 is recovered. This mixture is stable over a period of several days; it can also be spin cast and is UV curable. We successfully spin cast and patterned the SU-8/gold mixture into samples containing up to 14 vol% gold. The nanocomposite samples developed with well-defined edges, successfully released from the wafer, and no gold particles were observed to leech out of the developed samples.

The SU-8/gold nanoparticle mixtures were spin cast on 100-mm wafers and lithographically patterned into ASTM standard D638 tensile test specimen geometries. The samples were released by dissolving a sacrificial layer of MicroChem Omnicoat resist beneath the specimens.

Diamantane of varying wt% were physically mixed with SU-8. The crystals were ground by hand using a mortar and pestle, and mixed with the SU-8.

SWNTs are physically incorporated into the SU-8 photoresist. We were able to successfully spin cast wafers with SU-8/SWNT nanocomposite mixtures at 1 and 5 wt%, respectively. The SWNTs were dissolved in chloroform of equal volume and sonicated with the SU-8 for 1.5 days. The mixture was then spin cast onto an aluminum coated wafer and cured according to the manufacturer recommended procedures [2]. The devices were released from the substrate using an anodic dissolution method [21].

### 2.2. SU-8 and Nanocomposite Processing

As shown in Table 1, several formulations of SU-8 photoresist were used in the experiments. To spin cast, pattern and develop the SU-8/nanoparticle mixtures, MicroChem recommended procedures for 100-um (nominal) film thicknesses were followed. The typical procedure includes substrate pretreatment (rinse with dilute acid, DI water rinse), spin coat, soft bake, expose, post expose bake (PEB), develop, rinse & dry, and release of the devices. We used two different release methods: Omnicoat and anodic dissolution. For the Omnicoat release method, approximately 10-mL is placed in the center of a clean wafer. The wafer is accelerated at 100 rpm/s to a spin speed of 500 rpm for 5 sec. The spin speed is increased at 300 rpm/s to 3000 rpm for 30 sec. The wafer is then baked on a hotplate at 200-deg C for 1 minute. The SU-8/nanoparticle mixture is now applied to the wafer.

| Representative Test | SU-8 Type | Nanoparticle Type | % Particles | Mixing Method | Nominal Thickness (μm) | Release Method | Elastic Modulus (GPa) |
|---|---|---|---|---|---|---|---|
| Control | 50 | none | NA | NA | 100 | Anodic | 1.6 |
| Gold | 2035 | Gold | 14 vol | solvent | 110 | OmniCoat | 1.5 |
| D-Test 2 | 50 | Diamantane | 7 wt | stirring | 100 | Anodic | 1.9 |
| D-Test 5 | 50 | Diamantane | 5 wt | chloroform | 100 | Anodic | 1.4 |
| D-Test 6 | 50 | Diamantane | 1 wt | stirring | 100 | Anodic | 1.5 |
| D-Test 7 | 50 | Diamantane | 3 wt | stirring | 100 | Anodic | 1.6 |
| SWNT-Test 1 | 50 | SWNT | 0.25 wt | chloroform | 100 | Anodic | 1.3 |
| SWNT-Test 2 | 50 | SWNT | 1 wt | chloroform | 100 | Anodic | 0.54 |
| SWNT-Test 3 | 50 | SWNT | 5 wt | chloroform | 100 | Anodic | 0.26 |

**Table 1** Representative test data and experimentally derived elastic modulus.

The anodic dissolution release method [21] requires deposition of a layer of aluminum (500-nm thick) onto a silicon wafer. After the SU-8 mixture is spin cast and developed, the wafer is placed in a glass dish with water and salt crystals. 0.5-V is applied to the wafer and solution with the cathode attached to the wafer and the anode placed in the solution. The galvanic reaction removes the layer of aluminum and releases the devices. Typical removal time for these large devices is 16 hours. The anodic dissolution method is non-toxic.

| Lorenz [3] | 4 |
|---|---|
| Dellman [23] | 4.95 |
| Damean [13] | 3 |
| Hopcroft [22] | 2-3 |
| This study | 1.6 |

**Table 2** Reported moduli (GPa) for standard processing of SU-8.

To spin cast the nanocomposite mixture onto a wafer, approximately 10-mL is dispensed onto the center of the wafer. The wafer is then accelerated at 100 rpm/s for 10 seconds until the spin speed of 500 rpm is reached. The spin speed than increases at 300 rpm/s to 1000 rpm the run for an additional 30 seconds. The wafer is baked on a hot plate at 65-deg C for 5 minutes, then at 95-deg C for 20 minutes. An OAI collimated near-UV light source (350 to 450-nm wavelength) is used for device patterning. The devices are exposed to an intensity of 20mW/cm^2 for 2x11-sec exposures and patterned through a mask. After completing exposure, the wafer is baked at 65-deg C for 1 minute, then 95-deg C for 10 minutes. The devices are removed using either the Omnicoat or anodic dissolution method.

### 2.3. Mechanical and Electrical Properties

Uniaxial tensile tests and resistivity measurements were conducted on the ASTM D638 dogbone and rectangular nanocomposite shapes using a MTS Bionix 200 tensile tester and Interface SM-10 10-lb load cell. Our experimental setup uses custom conductive (brass) grips. The grips are connected to a Keithley 487 Picoammeter/Voltage source which interfaces with a LabView datalogger and allows setup of a voltage across the grips for 4-wire measurements of current. Resistance is inferred from voltage and current measurements and the gauge factor calculated. application via GPIB. A BNC cable attaches to either end of the grips  Viscosity measurements were performed using a Rheometric Scientific, Inc. ARES 3- LS. The 50-mm tool type has a nominal cone angle of 0.2 radians. 0.7-mL samples of control SU-8 and SU-8/diamantane mixtures were tested.

### 3. RESULTS AND DISCUSSION

#### 3.1. SU-8/Gold Nanocomposites

SU-8 has been successfully modified with suspensions of gold nanoparticles. Spherical gold particles were incorporated into the SU-8 polymer with the aim of making the resulting SU-8/gold nanocomposite conductive yet still processable by spin casting and lithography. Previous research using silver powders [10] resulted in high viscosity mixtures requiring the material to be spread onto the wafer.

Electrical conductivity tests showed undeformed resistivities greater than GΩ in samples with 14 vol % gold nanoparticles. Small, uncorrelated resistance changes were observed under strain. Percolation theory [15] suggests that for uniformly spherical nanoparticles and face-centered cubic (FCC) lattices, a volume fraction of 12% provides a continuous path for conduction, and this volume fraction is larger for other configurations, e.g. 25% for simple cubic lattices. This discrepancy could be due to non-uniform particle dispersion in the matrix. Our 14 vol% sample was the highest spin castable formulation.

As shown in Figure 3a, samples were observed to have lower residual stresses than control SU-8 samples spin cast and patterned with the same lithographic exposure, bake, and development process. This observation is consistent with reports by Renaud on SU-8 composites with silica nanoparticles [11].

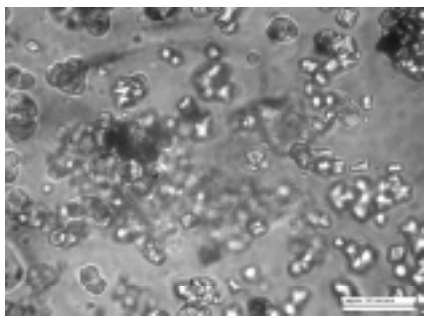

**Figure 2** Optical micrograph of SU-8/Diamondoid nanocomposite fabricated by physical incorporation of nanoparticles.

### 3.2. SU-8/Diamondoid Nanocomposites

Incorporation of diamantane in SU-8 provides the expected reduction in residual stress, as shown in Figure 3b. Diamantane were physically incorporated into the SU-8 and spin cast onto an aluminum coated silicon wafer.

Since diamantane can form nonreactive macroscopic crystals held together by van der Waals forces [19,20], several samples were sonicated with chloroform to improve mixing, but these attempts yielded similar results: the diamantane formed crystals of different sizes. (Figure 2). The non-toxic anodic dissolution method removed the sacrificial layer of aluminum [21]. Although the process is slower than the recommended Omnicoat release layer [2], it does not generate hazardous waste.

The preliminary residual stress results for the SU-8/diamantane specimens show a decrease compared to control SU-8 samples. Residual stress is calculated from beam theory

$$\sigma_R = \frac{Eth}{r_c}$$

where $E$ is the experimental effective modulus, $t$ and $h$ are geometric dimensions of the samples, and $r_c$ is the radius of curvature based on sample geometry.

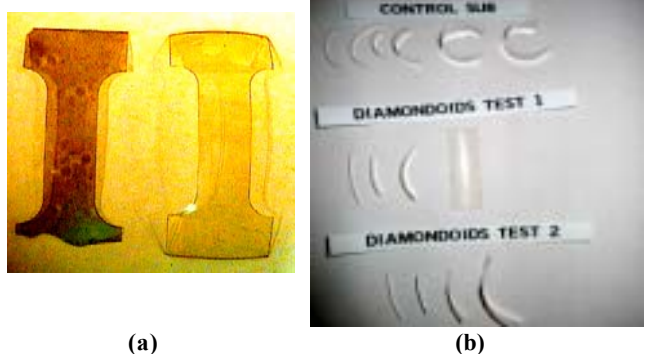

(a)                                  (b)

**Figure 3** a) ASTM standard D638 dogbone tensile test specimens. (left) SU-8/gold nanocomposite, (right) control SU-8. b) Comparison of residual stresses for SU-8/diamantane nanocomposites with control SU-8. Top: control SU-8; middle: 7 wt% SU-8/diamantane; bottom: 5 wt% SU-8/diamantane.

In the upper section of Figure 4, the reduction in residual stress is illustrated for the SU-8/diamantane samples. Table 1 shows the type of SU-8 used as well as information about the fabricated nanocomposite materials.

The SU-8/diamantane test specimen 11 is an ASTM standard D638 dogbone. The results of tensile tests are shown in Figure 5. The SU-8/gold nanocomposite elastic modulus (1.5 GPa) was extracted from a fit to the linear region of the stress-strain curve generated from tensile tests. This value is lower than moduli (Table 2) reported for SU-8 similarly processed using the manufacturer's specifications. If we consider the interaction between the matrix and the diamantane crystals as a large-particle composite, we can use the rule of mixtures to determine an upper and lower bound for the elastic modulus of the nanocomposite. The lowest value for the elastic modulus is our experimental elastic modulus for control SU-8: 1.7-GPa. As the wt% of diamantane increases, the elastic modulus should increase as the nanocomposite exhibits the properties of the diamantane. The elastic modulus for a diamantane molecule is expected to be high (on the order of diamond, 700-1200 GPa), but when diamantane crystallizes, the crystal is much "softer" due to hydrogen bonds [24]. Since the elastic modulus of the diamantane molecule is not well-defined, an upper bound for the SU-8/diamantane nanocomposite requires estimating this value: the diamantane elastic modulus is significantly higher than SU-8 (2 to 3 orders of magnitude higher) [24].

From our experimental results, the increased wt% of diamantane decreases the modulus below the lower bound established by the control SU-8. The expected increase in modulus, except for the 7 wt% mixture, does not correlate to the large particle composite increase in elastic modulus as filler increases in percent volume. This result contrasts previous estimates of an increased elastic modulus of SU-8 composites with the addition of nickel nanospheres by physical incorporation [13].

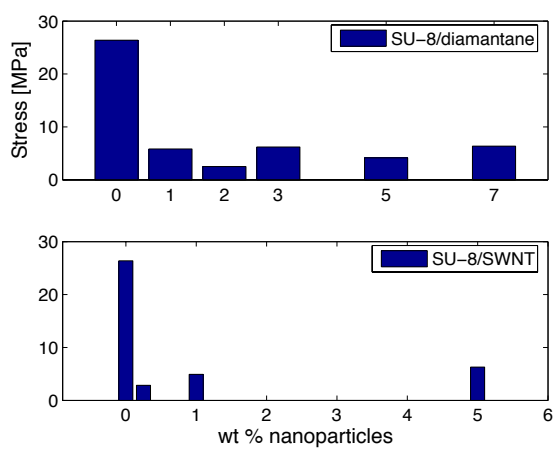

**Figure 4** Average residual stress results for SU-8/diamantane. The non-uniformity of specimen thickness imparts measurement errors up to 30%.

Figure 6 shows viscosity results for SU-8/diamantane tests at various wt%. Compared with the control SU-8, viscosity decreases with the addition of diamantane. The result for the 5 wt% SU-8/diamantane mixture varies from the other wt% samples, where experimental error or larger crystals may have contributed to the discrepancy. Future viscosity tests will address these repeatability issues.

### 3.3. SU-8/SWNT Nanocomposites

Our aim is to improve processing methods for repeatability and fabricating devices exhibiting decreased residual stresses and improved mechanical and electrical properties. Therefore, we also fabricated ASTM standard D638 test specimens of 0.25, 1, and 5 wt% of single wall carbon nanotubes (SWNTs). With physical incorporation, the SWNTs were unevenly dispersed in the SU-8. Additionally, we encountered difficulties in releasing the test specimens from the substrate using both release methods and as a result, these tensile test specimens were rectangular in shape rather than the dogbone geometry.

As shown in Figure 7, the elastic moduli of the SU-8/SWNT nanocomposites are less than the control SU-8. Resistivity measurements show a linear trend for the 1 wt% SU-8/SWNT nanocomposite (Figure 8) with a gauge factor of approximately 2 to 4. The 0.25 wt % SU-8/SWNT nancomposite was non-conductive, so resistivity results are shown for 1 and 5 wt% only. As shown in the lower section of Figure 4, residual stresses decreased for the SU-8/SWNT nanocomposites compared to the control SU-8 specimens.

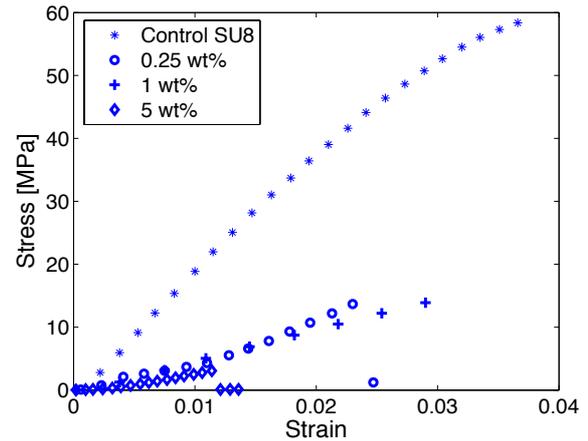
**Figure 7** SU-8/SWNT tensile test results.

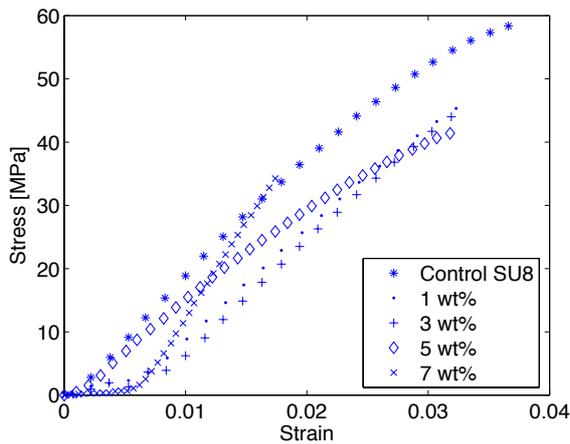
**Figure 5** SU-8/diamantane tensile test results: specimen 11

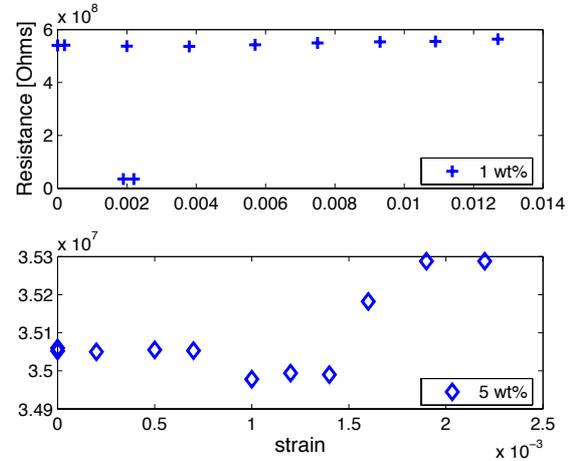
**Figure 8** Resistivity data for SU-8/SWNT specimen 8.

### 4. FUTURE WORK

Continuing research will optimize processing and integration of SU-8 with functionalized carbon nanotubes and diamondoids. Chemical modification of the nanoparticles should allow better control of dispersion and chemical bonding to the SU-8 matrix. This in turn may enable the ability to tailor mechanical and electrical transport properties.

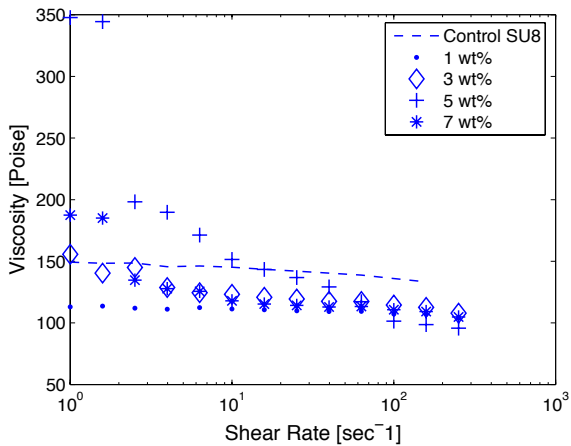
**Figure 6** Viscosity measurements of SU-8/diamantane. Repeatability/unifomity of spin coating may be affected.

## 5. CONCLUSIONS

We have shown spin-cast, lithographically curable SU-8/gold, SU-8/diamantane, and SU-8/SWNT nanocomposites for microfabricated structures with reduced residual stress and potentially tunable stiffness properties. We found the effective modulus of the SU-8 test specimens decreases with the addition of nanoparticles compared with control SU-8 values. We also observed decreasing viscosity with the addition of diamantane to SU-8. The SU-8/SWNT nanocomposites showed decreased elastic modulus compared to the control SU-8 as well as changes in resistivity due to increased strain. The gauge factor for the 1 wt% SU8/SWNT nanocomposite is approximately 2 to 4.

## ACKNOWLEDGEMENTS

This material is based upon work supported by the National Science Foundation under Grants No. EEC-0425914, ECS-0403769, and GRF (HC). The authors would like to thank the E. Shaqfeh Lab, R. Dauskardt group, Jeremy Dahl and David Tomanek.